\documentclass[oldversion,preprint]{aa}
\usepackage[varg]{txfonts}
\usepackage{graphicx}

\begin{document}

\title{Limit to the radio emission from a putative central compact source in SN1993J} 

\titlerunning{Limit to a putative central source in SN1993J}

\author{I. Mart\'i-Vidal\inst{1} 
\and 
J. M. Marcaide\inst{2}} 

\offprints{I. Mart\'i-Vidal, \email{mivan@chalmers.se}}

\institute{Onsala Space Observatory, Chalmers University of Technology, 
Observatoriev\"agen 90, SE-43992, Onsala, Sweden
\and 
Dept. Astronomia i Astrof\'isica, Universitat de Val\`encia, 
Dr. Moliner 50, ES-46100 Burjassot, Spain} 

\date{Accepted for publication in A\&A}

\abstract{SN\,1993J in M\,81 is the best studied young radio-luminous supernova in the Northern Hemisphere. We recently reported results from the analysis of a complete set of VLBI observations of this supernova at 1.7, 2.3, 5.0, and 8.4\,GHz, covering a time baseline of more than one decade. Those reported results were focused on the kinematics of the expanding shock, the particulars of its evolving non-thermal emission, the density profile of the circumstellar medium, and the evolving free-free opacity by the supernova ejecta. In the present paper, we complete our analysis by performing a search for any possible signal from a compact source (i.e., a stellar-mass black hole or a young pulsar nebula) at the center of the expanding shell. We have performed a stacking of all our VLBI images at each frequency, after subtraction of our best-fit shell model at each epoch, and measured the peak intensity in the stacked residual image. Given the large amount of available global VLBI observations, the stacking of all the residual images allows us to put upper limits to the eventual emission of a putative compact central source at the level of $\sim102$\,$\mu$Jy at 5\,GHz (or, more conservatively, $\sim192$\,$\mu$Jy, if we make a further correction for the ejecta opacity) and somewhat larger at other wavelengths.
} 

\maketitle

\section{Introduction}

SN\,1993J in the galaxy M\,81 is, by far, the best studied young radio-luminous supernova in the Northern Hemisphere. The high declination and moderate radio-loud AGN activity of the host galaxy, M\,81, allowed us to perform an intensive VLBI monitoring of the supernova with a high-quality phase calibration, using the closeby AGN in M\,81 as calibrator source. Results on the analysis of the VLBI data (expansion curve and evolution in the shell internal structure) can be found in, e. g., Marcaide et al. (\cite{Jon94}, \cite{Jon97}, \cite{Jon09}), Bartel et al. (\cite{Bartel02}), Bietenholz et al. (\cite{Bieten03}), and Mart\'i-Vidal et al. (\cite{PaperI}); the analysis and modelling of the complete set of multi-frequency radio lightcurves (from 0.3 to 86\,GHz) can be found in Weiler et al. (\cite{Weiler}); and, finally, a combined self-consistent analysis of both, VLBI observations and radio lightcurves (with implications for the CSM radial density profile and the absorption mechanisms at the shocked medium and thermal ejecta) can be found in Mart\'i-Vidal et al. (\cite{PaperII}).

The strong and spherical radio emission of SN\,1993J and the high quality of the VLBI observations, also allowed us to use the supernova shell as an astrometry reference to study the AGN in M\,81. Results on the core-jet structure of the AGN and its frequency core-shift can be found in Bietenholz et al. (\cite{BietenM81}) and Mart\'i-Vidal et al. (\cite{IMVM81}); in the latter, strong evidence of jet precession, coupled to changes in the flux density of the jet, have also been reported.

It is well known that the central object expected to result from a core-collapse supernova is either a black hole or a neutron star. In either case, radio emission may be produced, due to accretion (i.e., the case of a black hole) or to a young pulsar nebula, that would form around the neutron star. Hence, the search for compact radio emission in young core-collapse supernovae (not associated with the non-thermal synchrotron radiation produced at their expanding shocks) is of particular importance, in order to check directly the basic tenets of the current models. Detection of such compact radio sources would observationally relate either a stellar-mass black hole or a neutron star to a young core-collapse supernova.

A compact source with an inverted spectrum was discovered in supernova SN\,1986J from VLBI observations (Bietenholz et al. \cite{Bieten86JSci}). Such a compact source, roughly located at the projected centroid of the distorted radio shell, was then identified as the first likely candidate of radio emission from a black hole, or a young pulsar nebula, associated with a modern supernova. However, the late evolution of the spectrum of this component, which is now similar to the spectrum in the outer shell (Bietenholz et al. \cite{Bieten86JApJ}), makes an alternate scenario also plausible: this central source could indeed be related to the interaction of the expanding supernova shock with a strong condensation of circumstellar plasma, located by chance at the projected center of the shell. Hence, the ultimate nature of this intriguing central component in SN\,1986J remains unclear.   

Pulsar searches in young supernova remnants have been performed also in the optical, as it is the case of SN\,1987A (e.g., Middleditch et al. \cite{Mid2000}), but no detection has been been confirmed so far.

Regarding SN\,1993J, emission from a pulsar was suggested by Woosley et al. (\cite{Woosley}) and Shigeyama et al. (\cite{Shige}), but these claims were refuted by Marcaide et al. (\cite{Jon97}), who put a limit of 500\,$\mu$Jy to any central compact emission at 5\,GHz.

In the present publication, we complete the analysis of our set of SN\,1993J VLBI observations by performing a search for any possible signal from a compact source at the center of the expanding shell. We maximize the sensitivity of our study by performing a stacking analysis of all the available epochs when the shell could be resolved at each frequency. In Sect \ref{Observations}, we briefly describe our set of observations; in Sect. \ref{Stacking}, we describe the details of our stacking analysis; in Sect. \ref{Results}, we present our results; in Sect. \ref{Conclusions}, we summarize our conclusions.

\section{Observations and data analysis}
\label{Observations}

A detailed description of our complete VLBI dataset can be found in Mart\'i-Vidal et al. (\cite{PaperI}). The observing epochs are distributed in time from year 1993 to the end of year 2005. However, since we are interested in the detection of emission from the central region of the supernova shell, we can only use the subset of epochs when the shell was well resolved at each frequency. This is the only way to decouple robustly the emission from the shell and the emission from a putative compact source at its center. This condition limits our study to the epochs beginning on December 1995 at the frequencies of 5.0 and 8.4\,GHz (i.e., when the shell radius was $\sim2$\,mas, which is roughly twice the typical beam width at these frequencies) and to the epochs beginning on November 2001 at the frequency of 1.7\,GHz (i.e., when the shell was resolved in a similar way as in the 5\,GHz images in year 1996). This makes a total of 33 epochs at 5\,GHz, 10 epochs at 8.4\,GHz, and 5 epochs at 1.7\,GHz.

\subsection{Stacking analysis}
\label{Stacking}

In Mart\'i-Vidal et al. (\cite{PaperII}), we report a model of the SN\,1993J radio emission that explains, simultaneously, the expansion curve measured with VLBI at the different frequencies and all the light curves reported in Weiler et al. (\cite{Weiler}). In this model, an evolving opacity of the inner supernova ejecta is needed, in order to explain frequency effects discovered in the expansion curve (first reported in Marcaide et al. \cite{Jon09}) and an intriguing flattening in the supernova spectrum at late epochs (see Figs. 1 and 8 in Mart\'i-Vidal et al. \cite{PaperII}, respectively).

The evolution in the free-free ejecta opacity used in our modelling is shown in Fig. 7 (left) of Mart\'i-Vidal et al. (\cite{PaperII}). According to this model, on June 1997, the ejecta opacity began to slowly decrease in time at high frequencies, while remaining still high at low frequencies (this is, indeed, an expected evolution for the free-free opacity in an expanding medium). Then, the ejecta became completely transparent at high frequencies around the beginning of year 2000, while the opacity remained high at low frequencies.  

We have taken the model reported in Mart\'i-Vidal et al. (\cite{PaperII}) and subtracted it from the restored images of the supernova shell at each epoch and frequency, resulting in a set of residual images of the supernova. Prior to the subtraction of the model from each image, we aligned all the images among them, using the frequency- and time-dependent shifts in the reference, reported in Mart\'i-Vidal et al. (\cite{IMVM81}). The stacking of the images was then performed as the simple pixel-wise average of all the residual images at each frequency. Each image was weighted in the average according to the inverse of its rms, as measured far from the location of the supernova emission.

Taking into account the ejecta opacity is very important for an accurate estimate of the flux density of a putative source at the center of the shell. Hence, we have also computed, for completeness, the stacked residual images in a similar way as described above, but subtracting shell models with no opacity evolution in the ejecta (i.e., assuming optically-thick ejecta at all frequencies and times).

\section{Results and discussion}
\label{Results}

\begin{figure*}
\centering
\includegraphics[width=17cm]{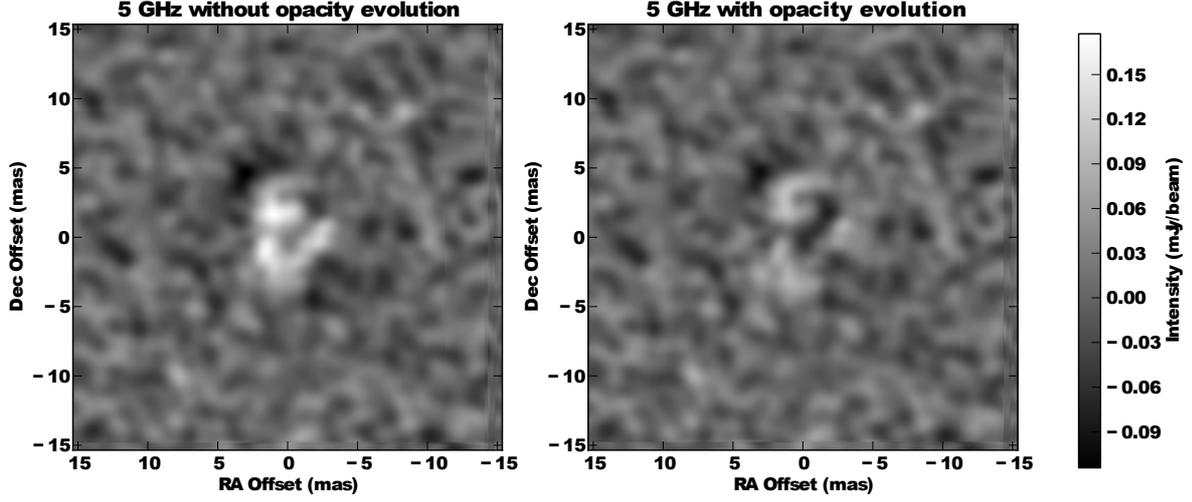}
\caption{Stacked residual images of SN\,1993J at 5\,GHz, using data from the end of year 1995 to the end of year 2005. Left, using a radio-emission model with constant opacity by the ejecta. Right, using the radio-emission model with evolving ejecta opacity reported in Mart\'i-Vidal et al. (\cite{PaperII}).}
\label{CStack}
\end{figure*}

\begin{figure*}
\centering
\includegraphics[width=17cm]{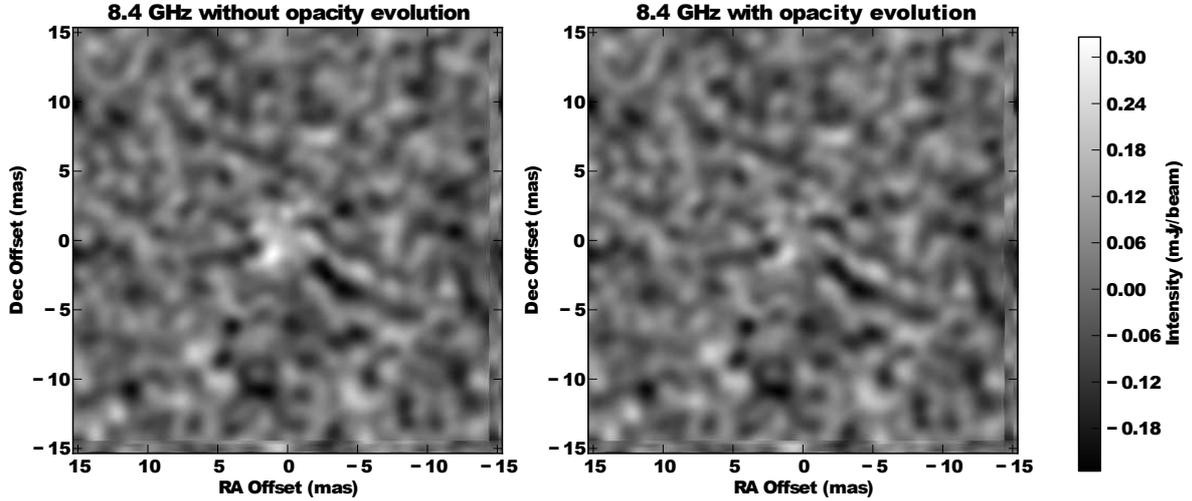}
\caption{Same as Fig. \ref{CStack}, but at 8.4\,GHz.}
\label{XStack}
\end{figure*}

\begin{figure}
\centering
\includegraphics[width=9.5cm]{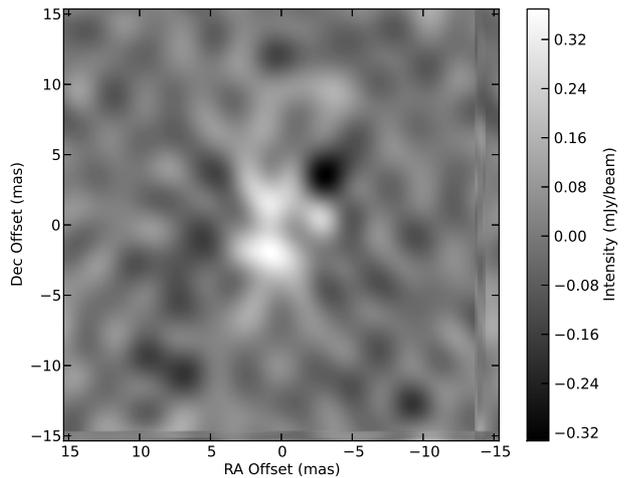}
\caption{Stacked residual image of SN\,1993J at 1.7\,GHz, using data from the end of year 2001 to the end of year 2005.}
\label{LStack}
\end{figure}

The stacked residual images are shown in Figs. \ref{CStack}, \ref{XStack}, and \ref{LStack} for the frequencies of 5, 8.4, and 1.7\,GHz, respectively. In the case of 5 and 8.4\,GHz, we show the images obtained from the subtraction of the shell model that takes opacity evolution into account (on the right side of the figures) and the subtraction of the model with constant opacity (on the left side of figures). The statistics for each image are given in Table \ref{TabRes}.

\begin{table*}
\caption{Statistics of the stacked VLBI residual images of SN\,1993J}
\label{TabRes}
\begin{center}
\begin{tabular}{c|cc|cc}\hline\hline
   & \multicolumn{2}{c|}{Without opacity evolution} & \multicolumn{2}{c}{With opacity evolution} \\
Freq. &  Peak & RMS & Peak & RMS \\
 (GHz) &($\mu$Jy\,beam$^{-1}$) & ($\mu$Jy\,beam$^{-1}$) & ($\mu$Jy\,beam$^{-1}$) &  (mJy\,beam$^{-1}$) \\\hline
  & & & & \\
1.7  & 370  & 70 & $-$  & $-$ \\
5.0  & 180  & 28 & 102  & 25 \\
8.4  & 320  & 68 & 245  & 66 \\\hline
\end{tabular}
\end{center}
{Note:}~The beam used for the intensity normalization is the average of the restoring beams in the stacked images.
\end{table*}

It can be readily seen in the figures that when the model with no opacity evolution is subtracted, the stacking brings clear residuals of positive intensity at the central region of the image (Figs. \ref{CStack} and \ref{XStack}, left). These residuals cannot be associated with a compact component, since they are spread through several beams. Hence, it is likely that they are produced by unmodelled contributions from the shell emission; this emission can indeed be well modelled if the opacity evolution reported in Mart\'i-Vidal et al (\cite{PaperII}) is taken into account (Figs. \ref{CStack} and \ref{XStack}, right). 

It can also be seen that, even taking the opacity evolution into consideration, the spatial distribution of the stacked residuals at 5\,GHz is not Gaussian at the central region of the image. This is likely due to the inhomogeneities in the supernova shell (see, e.g., Fig. 9 and Sect. 3.4.3 of Mart\'i-Vidal et al. \cite{PaperI}), which make it differ from a perfect spherically-symmetric model. These inhomogeneities map into blobs in the residual images at each epoch that add up into a unmodelled flux density in the stacked image. The effect of shell inhomogeneities on the stacked image is stronger at 1.7\,GHz, where large positive and negative peak residuals are obtained. We notice that, at this frequency, the supernova is well-resolved only at later epochs, when the total flux density is lower and the quality of the shell images is, thus, poorer.

\subsection{Limits to the emission by a central source}

The peak intensity in our stacked images is an upper-bound estimate to the emission of a central compact source in SN\,1993J. It is indeed a conservative estimate, since the peak is affected by the inhomogeneities in the shell, as discussed in the previous section. If we take into account the evolution in the ejecta opacity, and assume a constant flux density for an eventual source at the center of the shell, a still more conservative estimate of the upper-bound flux density, $S_{\mathrm{max}}$, is

\begin{equation}
S_{\mathrm{max}} = \frac{N\,S_p}{\sum_{i=1}^{N}{\left(1-\frac{\mathrm{Op}(t_i)}{100}\right)}},
\label{OpacCor}
\end{equation}

\noindent where $S_p$ is the peak intensity in the stacked image, $t_i$ is the observing time of the $i$-th epoch, $N$ is the number of epochs, and $\mathrm{Op}(t_i)$ is the percentage of intensity blocked by the ejecta at epoch $t_i$ (it is shown in Fig. 7, left, of Mart\'i-Vidal et al. \cite{PaperII}). The limiting flux density for a central source at 5\,GHz computed in this (more conservative) way increases to 192\,$\mu$Jy. At 8.4\,GHz, it increases to 1071\,$\mu$Jy. We notice that the value at 8.4\,GHz is high, because for the majority of the epochs at this frequency the putative source was observed when the opacity by the ejecta was high, according to the model in Mart\'i-Vidal et al. (\cite{PaperII}), so the real flux of the source (that would have been covered up by optically-thick ejecta most of the time) is much higher than the level of the residuals in the stacked image.

We also notice that were the flux density of a putative source rising with time, Eq. \ref{OpacCor} would further overestimate the flux-density upper limit.

\section{Conclusions}
\label{Conclusions}


We report upper-bound estimates to the emission of a putative compact radio source at the explosion center of supernova SN\,1993J. We have computed this upper bound as the peak emission in the image resulting from the stacking of all the residual images in the epochs when the shell structure was well-resolved by the interferometer at each frequency. By residual image, we mean the image resulting from the subtraction of the model reported in Mart\'i-Vidal et al. (\cite{PaperII}), which successfully explains, simultaneously, all the VLBI data and the complete radio lightcurves reported in Weiler et al. (\cite{Weiler}).

The upper bounds at 1.7, 5.0, and 8.4\,GHz are 370, 102, and 245\,$\mu$Jy, respectively.
These estimates do not include the correction for the free-free absorption by the inner supernova ejecta. More conservative estimates, which include those corrections, are 192 and 1071 $\mu$Jy at 5 and 8.4 GHz, respectively. At 1.7\,GHz, the ejecta are supposed to the opaque to the radio emission during the whole observing campaign. Hence, any emission from the shell center at this frequency would not be detected.

\begin{acknowledgements}

The National Radio Astronomy Observatory is a facility of the National Science Foundation operated under cooperative agreement by Associated Universities, Inc. The European VLBI Network is a joint facility of European, Chinese, South African, and other radio astronomy institutes funded by their national research councils. This research has been partially supported by projects AYA2009-13036-C02-01 and AYA2009-13036-C02-02 of the MICINN and by grant PROMETEO 104/2009 of the Generalitat Valenciana.

\end{acknowledgements}

\end{document}